\documentclass[preprints,article,accept,oneauthor,pdftex]{Definitions/mdpi} 

\firstpage{1} 
\pubvolume{1}
\issuenum{1}
\articlenumber{0}
\pubyear{2021}
\copyrightyear{2020}
\datereceived{} 
\dateaccepted{} 
\datepublished{} 
\hreflink{https://doi.org/} 
\pdfoutput=1

\Title{Generation of cold magnetized relativistic plasmas at the rear of thin foils irradiated by ultra-high-intensity laser pulses}

\TitleCitation{Generation of cold magnetized relativistic plasmas at the rear of thin foils irradiated by ultra-high-intensity laser pulses}

\Author{Artem V. Korzhimanov $^{1}$\orcidA{}}

\AuthorNames{Artem V. Korzhimanov}

\AuthorCitation{Korzhimanov, A. V.}

\address{%
$^{1}$ \quad Federal Research Center Institute of Applied Physics of the Russian Academy of Sciences; artem.korzhimanov@ipfran.ru}

\corres{Correspondence: artem.korzhimanov@ipfran.ru}

\abstract{A scheme to generate magnetized relativistic plasmas in laboratory is proposed. It is based on interaction of ultra-high-intensity sub-picosecond laser pulses with few-micron thick foils or films. By means of Particle-In-Cell simulations it is shown that energetic electrons produced by the laser and evacuated at the rear of the target trigger an expansion of the target and builds up a strong azimuthal magnetic field. It is shown that in the expanding plasma sheath a ratio of the magnetic pressure and the electron rest-mass energy density exceeds unity whereas a the plasma pressure is lower than the magnetic pressure and the electron gyroradius is lower than the plasma dimension. This scheme can be utilised to study astrophysical extreme phenomena such as relativistic magnetic reconnection in laboratory.}


\begin{document}

\section{Introduction}

Recently, there has been an increased interest in the possibility of investigating processes characteristic of extreme astrophysical objects in laboratories using powerful laser systems \cite{Remington2006, Bulanov2015a}. In particular, much attention is paid to the study of the magnetic reconnection process, which is widely encountered in space. Typical experiments in this area use two nanosecond laser pulses with energies ranging from a few Joules to several kilojoules and irradiate a metal surface. As a result of the interaction, ablation of a heated substance takes place, in which azimuthal magnetic fields are self-generated due to the Biermann battery effect. Such magnetized expanding plumes may be arranged to collide with each other, initiating a reconnection of oppositely directed magnetic lines in interaction region \cite{Nilson2006, Fox2011, Fiksel2014, Rosenberg2015, Matteucci2018}.

In most of the experiments carried out, however, the case of a nonrelativistic plasma was investigated. At the same time, magnetic reconnection in the relativistic regime are expected to occur in a vicinity of extreme astrophysical objects such as pulsars, magnetars, active galactic nuclei. A key parameter in this case is a cold magnetisation parameter, which is equal to a ratio of a magnetic pressure to an electron rest energy density $\sigma = B^2 / (\mu_0n_em_ec^2) $, where $B$ is the magnetic field induction, $\mu_0$ is the magnetic constant, $n_e$, $m_e$ are the concentration and mass of electrons, respectively, $c$ is the speed of light \cite{Lyubarsky2005}. The case of the magnetized relativistic plasma corresponds to $\sigma>1$, and until recently such values were unattainable in the laboratory.

In recent works, however, the $\sigma>1$ regime was obtained by two methods. One of them is similar to the approach of non-relativistic experiments, but it used either a pair of 2~J, 40~fs laser pulses from the HERCULES facility, or 500~J, 20~ps pulses from the OMEGA EP facility \cite{Raymond2018}. The high power of laser pulses made it possible to attain intensities of the order of $10^{18}$--$10^{19}$~W/cm$^2$ at the focus, which, as is known, is sufficient for the energy of the electrons to become comparable to their rest energy \cite{Mourou2006}. As a result, a field of about 1~kT was generated in the expanding plasma plume, which made it possible to reach $\sigma\sim10$. An integral property of the scheme used, however, was the high value of a parameter $\beta = \mu_0n_ekT_e/B^2$, equal to the ratio of the electron kinetic pressure to the magnetic pressure (here $k$ is the Boltzmann constant, $T_e$ is the electron temperature ). In the experiment, its value was estimated at $\beta\sim50$. At the same time, in the aspect of astrophysical applications, a cold plasma with a dominant role of a magnetic field and $\beta\ll 1 $ is usually considered.

The second method for achieving high degrees of magnetization in a laser-plasma experiment is based on the use of a picosecond laser pulse of a kilojoule energy level to generate a magnetic field in a microcoil \cite{Law2020}. When a coil was irradiated with laser radiation at opposite ends, counterstreaming currents of the order of 1~MA were generated in it, which lead to the creation of antiparallel fields of magnitude above 1~kT. According to estimates, the experiment reached $\sigma\sim20$--$100$ at $\beta\sim 0.01$--$0.1$. The main disadvantage of such experiments is a hard accessibility of such laser systems.

Other methods of organizing relativistic magnetic reconnection have been theoretically proposed as alternative approaches. For example, in the works \cite{Gu2015a,Gu2016b,Gu2016c,Gu2018,Gu2019,Gu2021} it is proposed to observe an antiparallel configuration of magnetic fields during a parallel propagation of two ultrashort superintense laser pulses in transparent plasma. In plasma wakefields generated by them, the azimuthal magnetic field is observed. When two wakefields come into contact, for example, when they gradually expand in plasma with a negative density gradient, those fields begin to annihilate. The disadvantage of this method is that the electrons in the interaction region are relativistic, and their Larmour radii exceed a size of the region of localization of magnetic fields. That is the reconnection is observed only in the electron diffusion region in the very vicinity of an X-point.

Finally, in a paper \cite{Yi2018} authors propose to use micro-slabs irradiated with laser pulses along a surface to observe relativistic magnetic reconnection. It was shown numerically that when using a femtosecond laser pulse of multi-terawatt power, $\sigma\sim 1$ can be achieved at $\beta\sim 0.1$, and when the power is increased to a petawatt level, $\sigma>100$ can be achieved. The main difficulty in the experimental implementation of such a scheme is the need to use high-contrast laser pulses and precise targeting of a relatively tightly focused pulse to the end of a slab of submicron thickness.

In this paper, we propose another method for generating cold ($\beta\sim 0.1$) magnetized ($\sigma\sim 10$) relativistic plasma, which is based on the interaction of powerful subpicosecond pulses with thin targets. As is known, the interaction of relativistically intense radiation with the surface leads to efficient generation of beams of energetic electrons injected by the field into the target \cite{Brunel1988, Mulser2012a, Liseykina2015}. These electrons, possessing relativistic energies, freely pass through targets up to several tens of microns thick. Escaping from the back side, they are decelerated by an emerging charge separation field and initiate ionization of the rear surface of the target and the expansion of the resulting plasma. As a result, a plasma sheath is formed, with electrons heated to relativistic temperatures and a laminar flow of cold ions \cite{Wilks2001}.

Continuing escaping electron bunches form a current on the axis of the sheath and generate an azimuthal magnetic field \cite{Robinson2009a}. At low laser pulse intensities and durations, this field does not significantly alter the expansion dynamics, although it can lead to the observed changes in the spectrum of the expanding ions. Recently, however, it was discovered that for pulses of several hundred femtoseconds duration and intensity above $10^{20}$~W/cm$^2$, the magnetic field is sufficient to magnetize the electrons and thereby appreciably slow down the expansion of the plasma \cite{Nakatsutsumi2018, Huang2021a}.

In this work, we studied in more detail the properties of the resulting plasma and found that conditions for a cold magnetized relativistic plasma are realized in it, which makes it possible to use this mechanism for observing relativistic magnetic reconnection under laboratory conditions.

\section{Methods}

An analysis of a formation of the magnetized plasma during the interaction of laser radiation with thin dense targets was carried out by numerical simulation by a particle-in-cell method using the PICADOR software package \cite{Surmin2016a} . Modeling was carried out in two-dimensional geometry. The length of the box was selected in the range from 280 to 380~$\mu$m, depending on the intensity and duration of the laser pulse, so that by the end of the simulation the plasma did not reach the right boundary. The box width was 80~$\mu$m. A grid step along both axes was 0.02~$\mu$m. A calculation time was 3000 fs, and a time step was 0.02~fs. Absorbing boundary conditions were used at all boundaries both for particles and fields. The number of particles in the cell at the initial moment of time was 200.

An aluminum foil with a thickness of 2~$\mu$m was used as the target. It was located at a distance of 30~$\mu$m from the left boundary of the box. Ions were preionized to the Al$^{9+}$ state. Their concentration at the initial moment of time was $5\times 10^{22}$~cm$^{- 3}$. The boundaries of the target were initially assumed to be perfectly sharp. A layer of fully ionized hydrogen with a thickness of 0.02~$\mu$m was added to the rear surface of the target, which imitated a natural contamination of the target surface, and also accelerated the expansion of the plasma due to the smaller mass of protons compared to the mass of aluminum ions.

A p-polarized laser pulse was incident on the target from the left along the normal to the surface and focused into a spot 4~$\mu$m in diameter at FWHM (Full Width at Half Maximum). The paper presents the results for pulses with a duration of 700~fs at FWHM; however, calculations were also performed with pulses with a duration of 100, 400, and 1000~fs. The results for pulses with the duration of 400 fs and longer are in qualitative agreement with each other, while for pulses with the duration of 100~fs the plasma stayed unmagnetized in the investigated range of intensities up to $10^{21}$~W/cm$^2$. The pulse envelope in both coordinates was a Gaussian.

The noisy distributions of electron density, electron energy density and magnetic field obtained as a result of modeling were smoothed by a Gaussian filter with a width of 0.2 ~ $\mu$m. All transformations have then been carried out on smoothed distributions. To obtain a magnetic field gradient, we used a convolution of the initial distribution of the magnetic field with the first derivative of the Gaussian function in both directions with the same width of 0.2~$\mu$m.

\section{Results}

\begin{figure}[H]
\includegraphics[width=13cm]{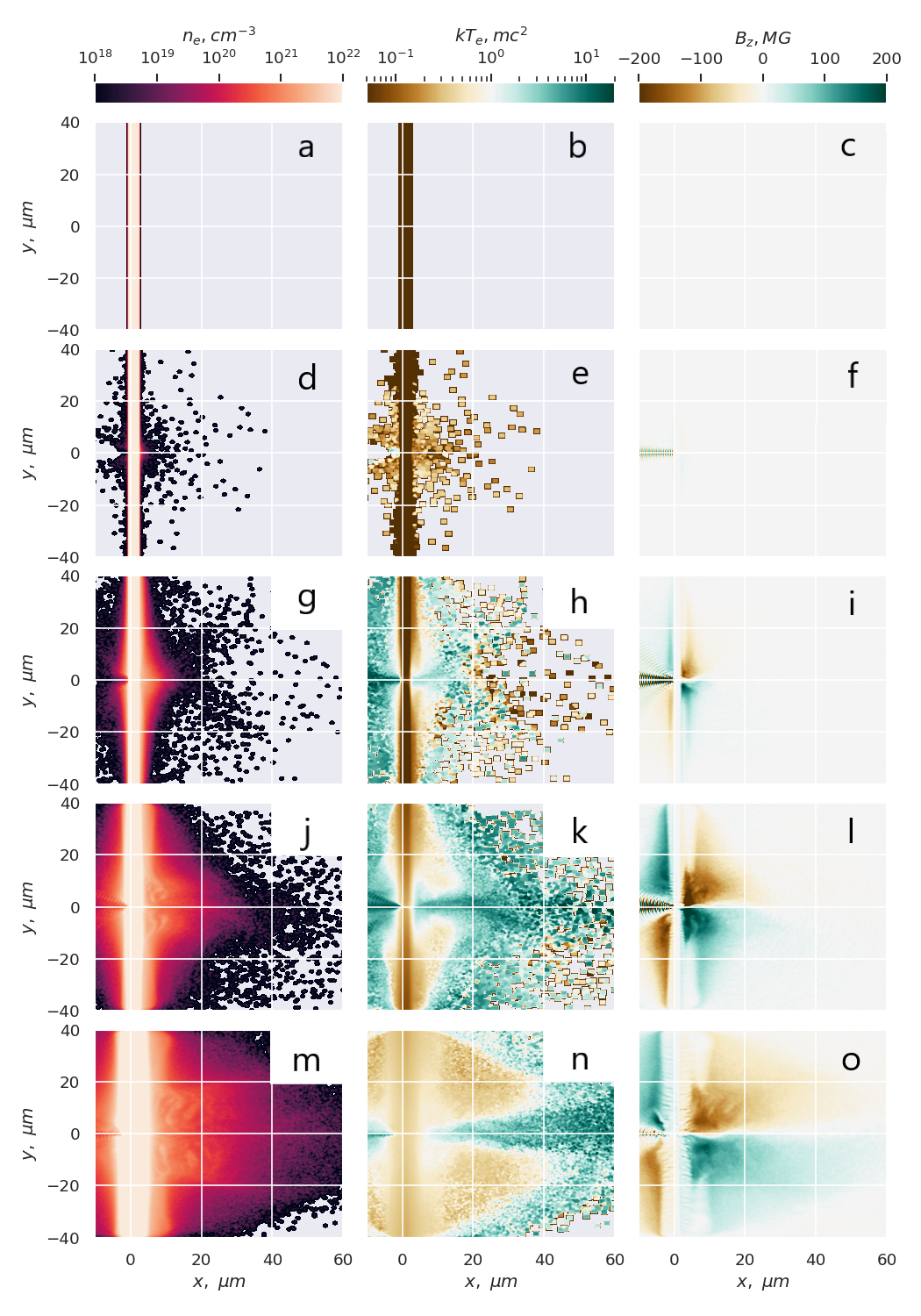}
\caption{Distributions of electron concentration (left column), electron temperature (central column) and transverse magnetic field (right column) at (a)-(c) $t=0$~fs, (d)-(f) $t=500$~fs, (g)-(i) $t=1000$~fs, (j)-(l) $t=1500$~fs, (m)-(o) $t=2000$~fs after start of the simulation. The parameters of the simulation are in the text.\label{fig:vs-time-1}}
\end{figure}

Fig.~\ref{fig:vs-time-1} shows the time evolution of the electron density $n_e$, the electron temperature $T_e$ (obtained as a ratio of the local kinetic energy density of electrons to their concentration) and the transverse component of the magnetic field $B_z$ at different times for a pulse with a duration of 700~fs and an intensity of $10^{20}$~W/cm$^2$.

It can be seen that at the beginning (Fig.~\ref{fig:vs-time-1} (d,e)) the target is deformed under the action of radiation and a noticeable pre-plasma layer is formed in the region $x<0$~$\mu$m, which enhances absorption of radiation. On the rear side of the target, the expulsion of the energetic electrons is observed. It triggers the expansion of the plasma and the formation of the plasma sheath. At a later moment in time (Fig.~\ref{fig:vs-time-1} (g)--(i)), an increase in the azimuthal magnetic field is observed in the formed plasma sheath. With increasing intensity, the magnitude of the generated magnetic field also grows, reaching a maximum at the moment when the maximum of the laser pulse arrives at the target (Fig.~\ref{fig:vs-time-1} (j)--(l)).

At this moment, the magnetic field at its peak reaches a value of the order of 400~MG, and is of the order of 100~MG in most of the plasma sheath. Note that the electron concentration in the same region is in the range $10^{20}$--$10^{21}$~cm$^{-3}$. As for the electron temperature, a high-energy electron current is observed on the sheath axis; however, to the side of it, the electron temperature is much lower and in a significant region near the target it is significantly below the relativistic value $kT_e<mc^2$. Due to the fact that in such a plasma there are no effective mechanisms for cooling electrons, then, most likely, these cold electrons were drawn out by a cold reverse current to compensate for the charge of hot electrons that had sufficient energy to leave the interaction region.

\begin{figure}[H]
\includegraphics[width=10cm]{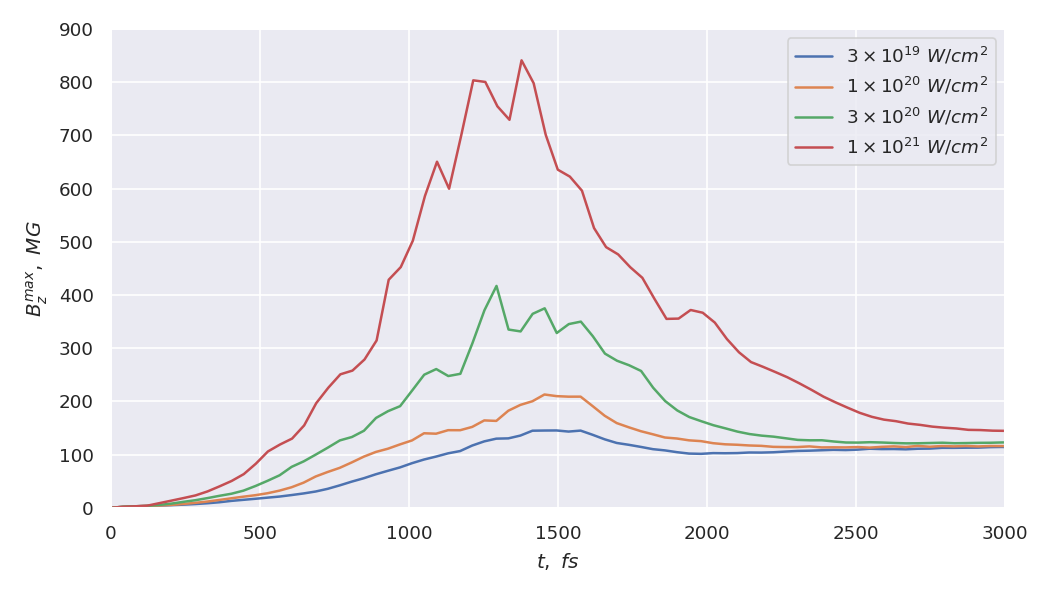}
\caption{Time evolution of the peak magnetic field generated in the expanding plasma sheath for different laser intensities.\label{fig:bz-vs-time}}
\end{figure}

At later times, the plasma continues to expand and cool down a little, but the magnetic field is dissipating slowly and changes in magnitude insignificantly. An interesting feature is that, as follows from the performed calculations, the magnitude of the magnetic field at later times is practically independent of the intensity of the laser radiation. This can be seen from Fig.~\ref{fig:bz-vs-time}, which shows a time dependence of the peak value of the magnetic field for different intensities of the laser pulses with fixed duration. Although the field value at the maximum of the laser pulse increases with intensity, it relaxes to an almost identical value of about 100~MG after the pulse leaves.

\begin{figure}[H]
\includegraphics[width=13cm]{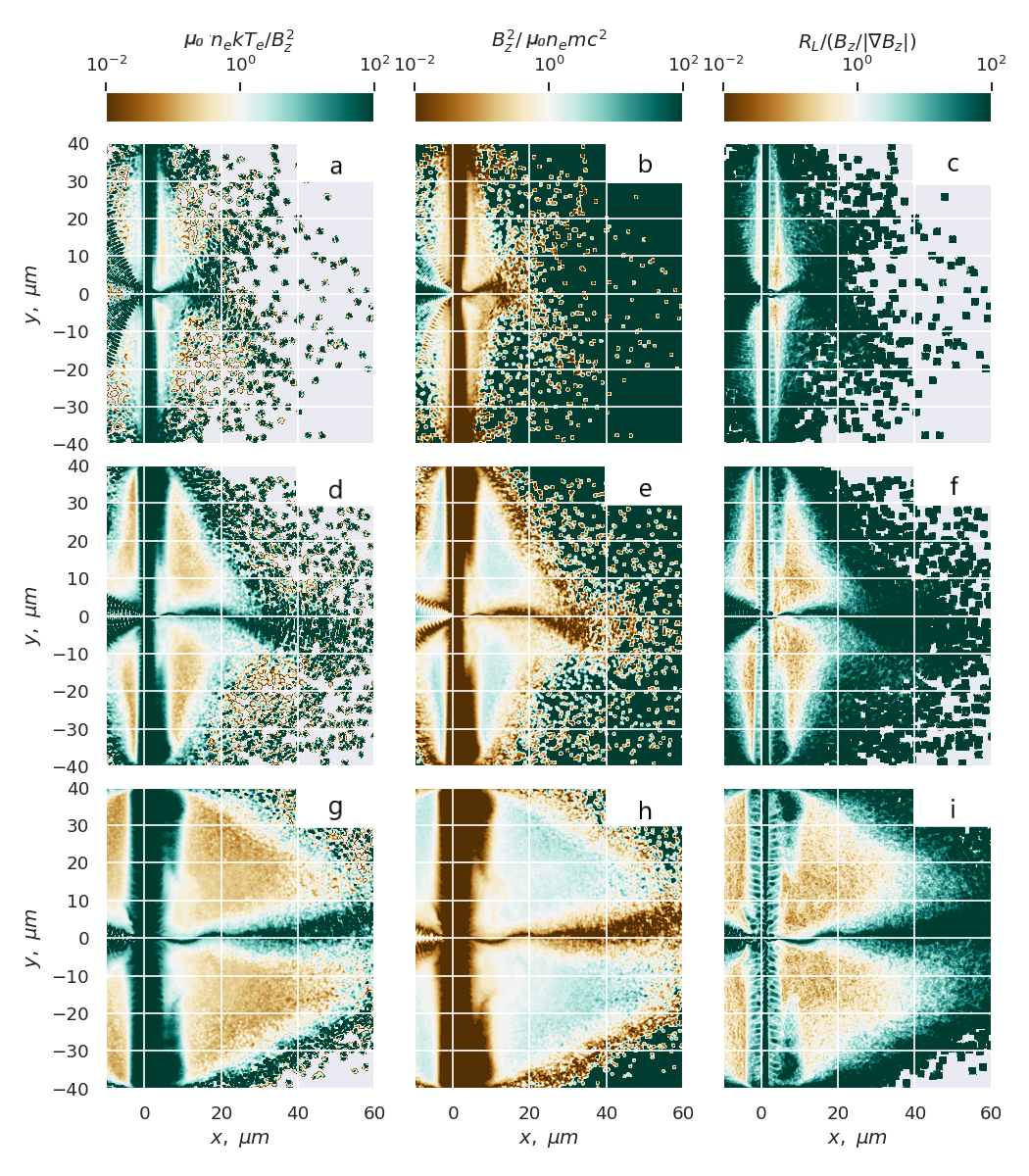}
\caption{Distributions of $\beta$ parameter (left column), $\sigma$ parameter (central column) and $\delta$ parameter (right column) at (a)-(c) $t=1000$~fs, (d)-(f) $t=1500$~fs, (g)-(i) $t=2000$~fs after start of the simulation. The parameters of the simulation are the same as at Fig.~\ref{fig:vs-time-1}.\label{fig:vs-time-2}}
\end{figure}

Let now turn to an analysis of the main parameters of the resulting plasma. Fig.~\ref{fig:vs-time-2} shows the distributions of the previously introduced parameters $\beta = \mu_0n_ekT_e/B_z^2$ and $\sigma = B_z^2/\mu_0n_em_ec^2$, and also a quantity $\delta = R_L / (B_z / |\nabla B_z|)$, equal to the ratio of the Larmour radius $R_L = \sqrt{kT_e/m_eB_z^2}$ to a characteristic scale of the magnetic field variation $L_B = B_z / |\nabla B_z|$. The analysis of $\delta$ is important because for effective plasma magnetization it is not enough to have a sufficiently large magnetic field, but it is also required that the electrons can make a full revolution in the magnetic field before it leaves the region of its localization. This can be a limiting factor in femtosecond laser plasma due to the small interaction volume and high gradients of the generated magnetic field. In particular, this is one of the limiting factors for the generation of magnetized plasma by pulses with a duration of less than 100~fs.

Within the discussed range of parameters, however, such a problem does not arise, although, as can be seen from Fig.~\ref{fig:vs-time-2}, the region in which $\delta<1$ turns out to be remarkably smaller than the region in which $\sigma>1$ and $\beta<1$.

Besides of that, it can be seen that at the beginning of the interaction, the plasma is non-magnetized (Fig.~\ref{fig:vs-time-2} (b)). However, by the arrival of the radiation maximum, a noticeable region of cold magnetized plasma appears (Fig.~\ref{fig:vs-time-2} (d)--(e)), which further begins to expand (Fig.~\ref{fig:vs-time-2} (g)--(h)). The achievable values of $\sigma$ are relatively small and do not exceed $10$.

\begin{figure}[H]
\includegraphics[width=13cm]{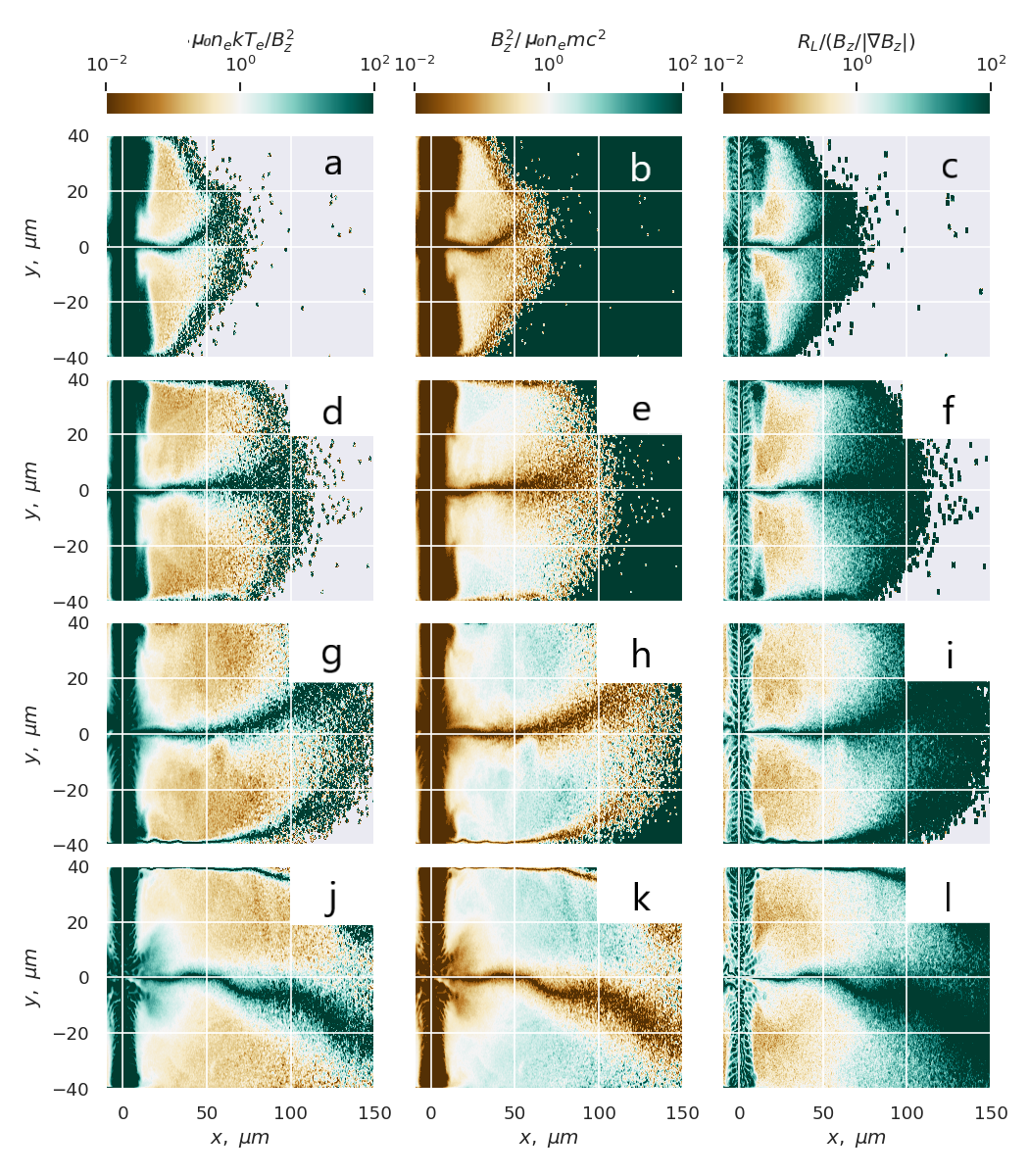}
\caption{The same as at the Fig.~\ref{fig:vs-time-2} but at $t=2500$~fs after start of the simulations for laser intensities of (a)-(c) $3\times10^{19}$~W/cm$^2$, (d)-(f) $10^{20}$~W/cm$^2$, (g)-(i) $3\times10^{20}$~W/cm$^2$, (j)-(l) $10^{21}$~W/cm$^2$.\label{fig:vs-intensity}}
\end{figure}

Achievement of larger values can be expected at higher intensities, therefore, we performed calculations in the range of intensities $3\times 10^{19}$--$10^{21}$~W/cm$^2$. The results are shown in Fig.~\ref{fig:vs-intensity}. It can be seen that at low intensities (Fig.~\ref{fig:vs-intensity} (a)--(c)) no magnetized plasma is formed. Although there is a noticeable region of low electron temperature in the sheath, the magnetic field is too weak to exceed the rest energy density of the electrons. With an increase in intensity, the plasma expanding with higher and higher speeds, which is in accordance with the theory of collisionless plasma expansion into vacuum \cite{Gurevich1966a, Mora1979a}. The magnitude of the generated magnetic fields also increases, and a significant region of cold magnetized plasma is formed. In this case, the degree of magnetization also increases and reaches $\sigma\sim 10$ for $10^{21}$~W/cm$^2$. Simultaneously, the value of $\delta$ remains small.

\section{Discussion}

Concluding, our numerical simulation showed that as a result of irradiation of thin (about a few microns thick) foils by relativistically intense laser pulses with a duration of several hundred femtoseconds, an expanding plasma sheath is formed on the rear side of the target, the conditions in which correspond to a cold magnetized plasma with parameters $\beta\sim 0.01$--$0.1$ and $\sigma\sim 1$--$10$. To create such conditions, an intensity higher than $10^{20}$~W/cm$^2$ is required, which is already available in experiments. To organize magnetic reconnection, two pulses are required, which are focused at a distance of about 50--100~$\mu$m from each other. This opens up new possibilities for the study of relativistic magnetic reconnection in the laboratory.

Additionally, we note that it can be expected that next-generation laser facilities carrying hundreds of joules in subpicosecond pulses will be able to create conditions in which the effects of radiation cooling and generation of electron-positron pairs become observable. Recently, it has been proposed that they can play an important role in relativistic magnetic reconnection \cite{Hakobyan2019, Hakobyan2021}. The study of such a possibility is an interesting possible subject of a future work.

\funding{This research was supported by Center of Excellence ”Center of Photonics” funded by The Ministry of Science and Higher Education of the Russian Federation, contract 075-15-2020-906. The simulations were performed on resources provided by the Joint Supercomputer Center of the Russian Academy of Sciences.}

\dataavailability{The data that support the findings of this study are available from the corresponding author upon reasonable request.}

\conflictsofinterest{The author declares no conflict of interest.}

\end{paracol}
\reftitle{References}


\externalbibliography{yes}
\bibliography{mag-rel-plas}

\end{document}